\documentstyle[11pt,epsfig]{article}
\newcommand{\ffig}[4]{\begin{figure}[hbt]\vfill\begin{center}
            \mbox{\epsfig{figure=#1,height=#2}}\caption{#3}\label{#4}
            \end{center}\vfill\end{figure}}
\title{\normalsize\bf  Note about a second "evidence" for a WIMP annual 
modulation}
\author{ \normalsize G. Gerbier$^a$, J. Mallet$^a$, L. Mosca$^a$, C. Tao$^b$}
\date{}
\begin{document}
\maketitle

\begin{flushleft}
{\footnotesize\it $^a$ DSM/DAPNIA/SPP, C.E.A. Saclay, F-91191 Gif-sur-Yvette,
France}

{\footnotesize\it $^b$ CPPM,IN2P3/CNRS and Universit\'e Aix-Marseille II, 163 
av. de Luminy, case 907, F-13288 Marseille cedex 09, France}

\end{flushleft}

\section {Introduction}
\hspace{0.5cm}
This note is intended to contribute to a clarification about a
claimed "evidence" by the DAMA group of an annual modulation of the counting
rate of a Dark Matter NaI(Tl) detector as due to a neutralino (SUSY-LSP) 
Dark Matter candidate [1]. A first "evidence" [2] had already given rise to 
a note of comments [3].

As the information given in this paper [1] refers only to the result of a 
theoretically constrained  fit, it is difficult to estimate the relevance of
the claimed evidence.

Answers to the following 5 questions would be essential to enable the 
scientific
community to correctly appreciate the relevance of the claimed effect to the 
search for WIMPs, which has been under way for several years, in a number of 
experiments over the world. 

\section {Questions}
\hspace{0.5cm} 1)  What are the experimental values, and the corresponding
experimental errors 
of the modulation amplitude Sm ?  This amplitude can be calculated from 
the daily counting rates [4], without any fit or maximum-likelihood 
procedure,
for each energy interval. 
The experimental errors should include both statistical and 
systematical contributions and be given separately.
Statistical errors for a given exposure can be 
calculated with good aproximation by anybody : in this case they are 
definitely larger than the errors from the maximum likelihood procedure 
presented in the last DAMA paper (see figure 1 of this note).

How can the strong difference between the distributions of the 
modulation amplitudes Sm in the two papers [1,2] be explained (see figure 2
of this note)? The only difference between the two approaches is the explicit 
presence of the WIMP
hypothesis in the maximum likelihood fit giving the Sm values in the second 
paper,
while in the previous paper the Sm values were obtained before introducing
this theoretical hypothesis.\\

2)  What is  the experimental distribution of the modulation amplitude Sm 
for the individual  NaI modules ? We remind that for the first data set (1/3 
of the new statistics), this distribution, given in one of the DAMA reports 
[2], was very unlike a "physical" distribution 
(there was an effect for 3 detectors while the 6 others did not show any 
deviation), as stressed in a note of comments [3] about this paper.\\

\ffig{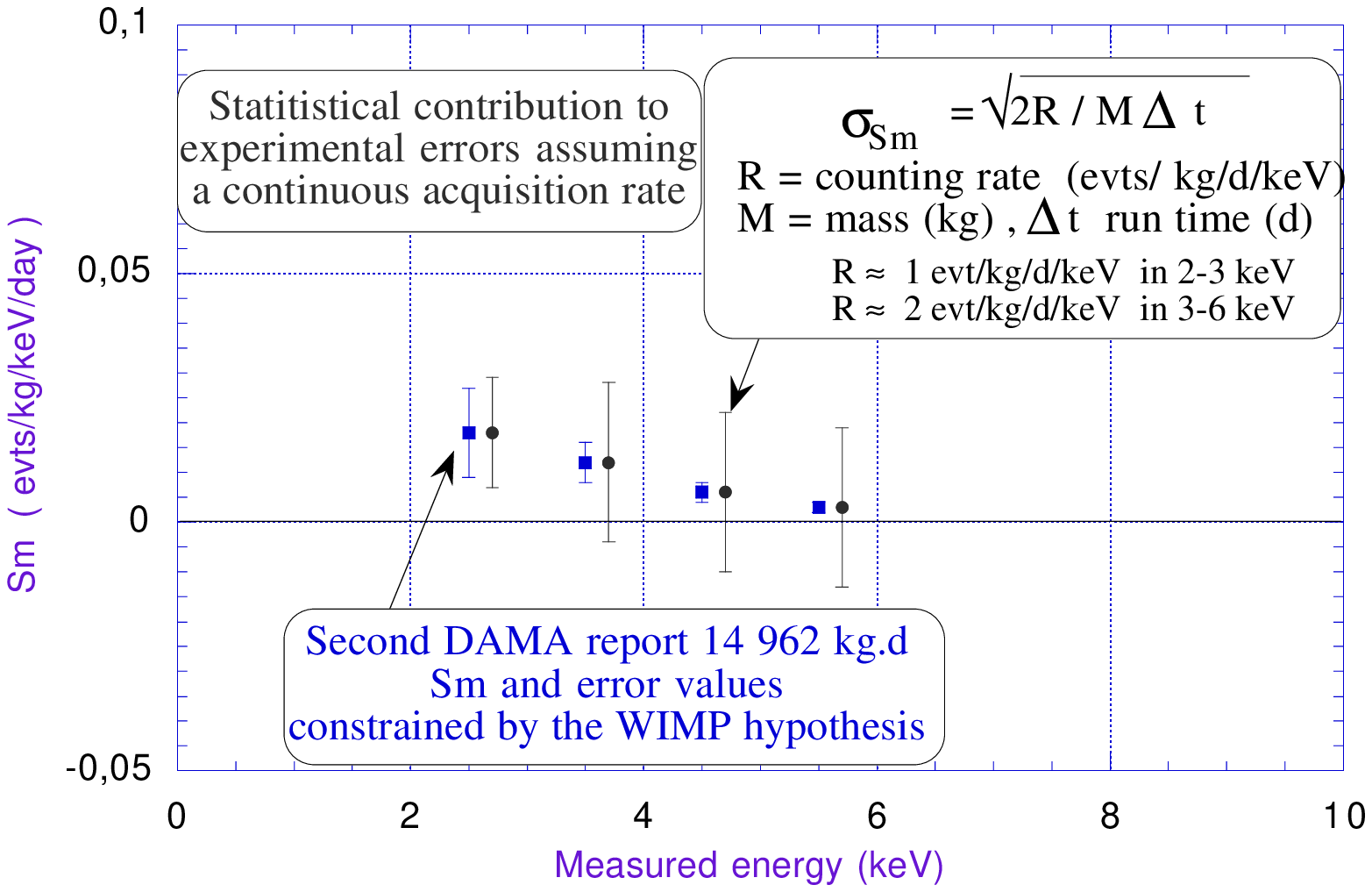}{80mm}{\it 
Modulation amplitude Sm distribution from the second exposure [1], 
constrained by the WIMP hypothesis. The corresponding  
errors coming from this DAMA fitting procedure are 
compared to the statistical contribution to the experimental errors, 
calculated by us, assuming a continuous acquisition. }{figuremd}

\ffig{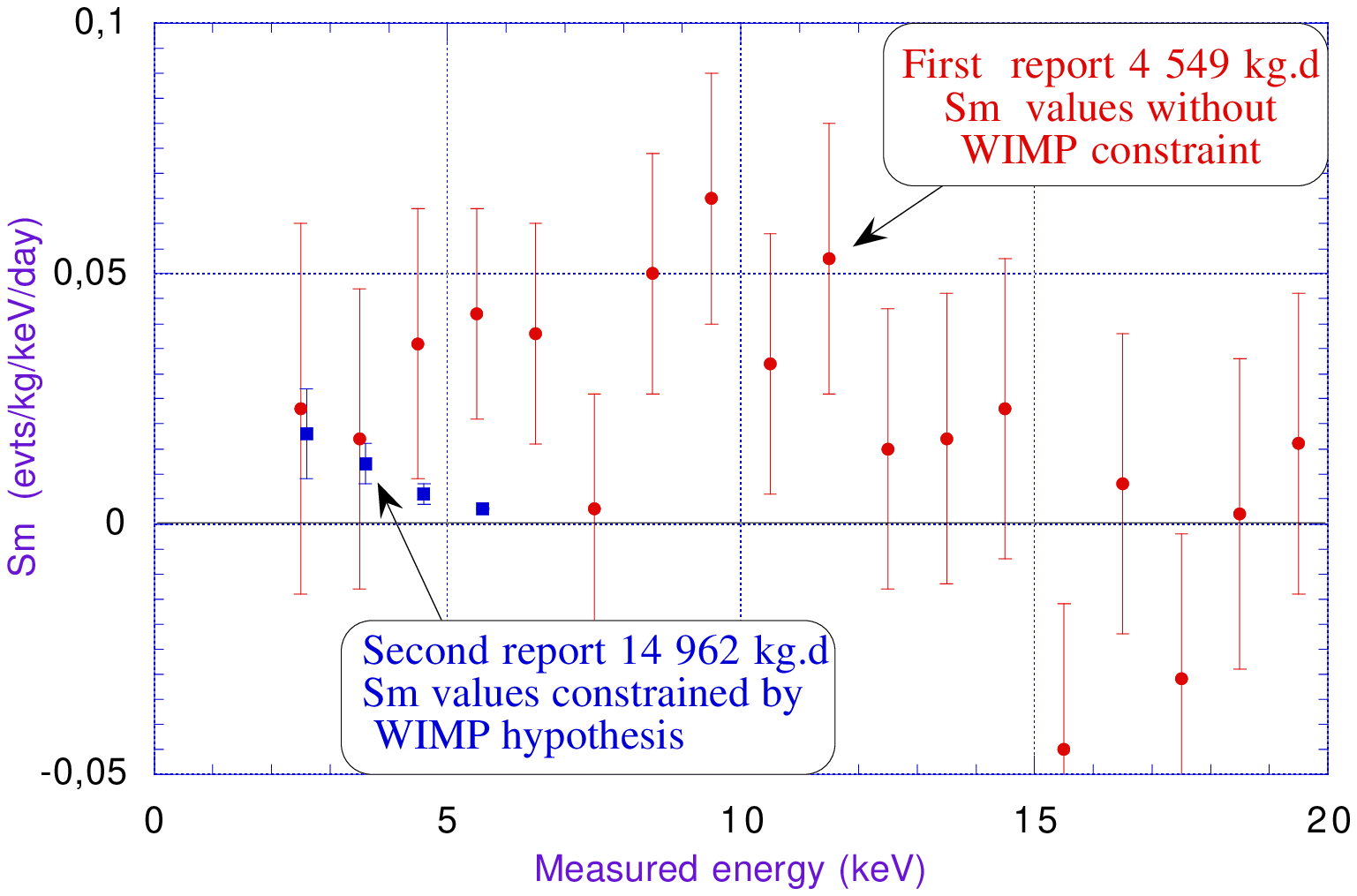}{80mm}{\it 
Modulation amplitude Sm values,constrained
by the WIMP hypothesis, from the second exposure [1] are compared 
with the Sm values from the previous paper, not constrained by the WIMP
hypothesis [2].}{figuremd}

3) Concerning possible systematic effects, can the "separation-plot" between 
physical 
events and PMT noise be shown for the 2-6 keV energy interval relevant to the 
claimed 
effect (and not for the large 2-20 keV interval as in figure 1 of ref [1] or 
figure 16 of ref [5])?

How can the strange behaviour of the 2-6 keV energy distributions of the nine 
crystals 
(figure 2 of ref [1]) be explained? The total counting rate in the 2-3 keV 
bin is typically a factor 2-3 smaller than the counting rates in the 
following bins. How can 
the even more important drop of the residual background (obtained after 
subtraction 
of the "signal" coming from the maximum likelihood fit) be accounted for?\\
2-3 keV  :    0.5 evts/keV/kg/day\\
3-4 keV  :    1.8 evts/keV/kg/day\\
4-5 keV  :    1.9 evts/keV/kg/day\\
5-6 keV  :    2.0 evts/keV/kg/day\\

4) The 2-6 keV region is affected by PMT noise which is partially rejected 
by software cuts. How is  the stability of the efficiency correction and 
contamination level proven to be not worse than 1$\%$, as absolutely needed 
for 
the claimed effect, in a region where the efficiency is strongly varying with 
energy 
(figure 17 of ref [5]) ?\\

5) In the presented maximum-likelihood procedure, which gives as output 
simultaneously an "evidence" for a rate modulation and the values of the two 
basic parameters caracterising  the neutralino candidate (mass and cross 
section on proton), what is the influence of the applied constraint on 
the mass (neutralino 
mass > 25 GeV) ?  In other words, if this constraint is removed, does at 
least one additional solution show-up in this maximum likelihood fit ?

\vspace{1cm}
\begin{large}
{\bf References}\\
\end{large}
\noindent  1.  DAMA collaboration, INFN/AE/98/20, ROM2F/98/34 \\
  2. 	R. Bernabei et al., ROM2F/97/33, Phys. Lett. B 424 (1998) 195-201 \\
  3. 	G. Gerbier et al., astro-ph/9710181\\
  4.	K. Freese, J. Frieman and A. Gould, Phys Rev D37 (1988)3388  \\ 
  5.  DAMA collaboration, INFN/AE/98/23 and ROM2F/98/27 \\ 
\end{document}